\documentclass[reqno,11pt]{amsart}
\usepackage{graphicx}
\usepackage{amscd}
\usepackage{slashed}
\usepackage{amssymb}
\usepackage{ushort}
\usepackage{mathtools}
\usepackage{esint}
\usepackage{pst-grad} 
\usepackage{pst-plot} 
\usepackage[mathscr]{eucal}
\numberwithin{equation}{section}
\allowdisplaybreaks[1]

\title[$L^p$-Spectrum of the Inverted Harmonic Oscillator]
{$L^p$-Spectrum of the Schr\"odinger Operator with Inverted Harmonic Oscillator Potential}

\author[F.\ Finster]{Felix Finster}
\address{Fakult\"at f\"ur Mathematik \\ Universit\"at Regensburg \\ D-93040 Regensburg \\ Germany}
\email{finster@ur.de}

\author[J.M.\ Isidro]{Jos{\'e} M. Isidro \\ \\ July 2017}
\address{Instituto Universitario de Matem\'atica Pura y Aplicada \\ Universidad Polit\'ecnica de Valencia \\
Valencia 46022 \\ Spain}
\email{joissan@mat.upv.es}

\newtheorem{Def}{Def.}[section]
\newtheorem{Thm}[Def]{Theorem}
\newtheorem{Prp}[Def]{Proposition}
\newtheorem{Lemma}[Def]{Lemma}

\newcommand{\Thanks}{\vspace*{.5em} \noindent \thanks}
\newcommand{\Proof}{\begin{proof}}
\newcommand{\QED}{\end{proof} \noindent}

\newcommand{\C}{\mathbb{C}}
\newcommand{\R}{\mathbb{R}}

\newcommand{\N}{\mathbb{N}}

\newcommand{\beq}{\begin{equation}}
\newcommand{\eeq}{\end{equation}}

\renewcommand{\O}{\mathscr{O}}

\DeclareMathOperator{\im}{Im}

\begin{document}
\maketitle

\begin{abstract}
We determine the $L^p$-spectrum of the Schr\"odinger operator
with the inverted harmonic oscillator potential $V(x)=-x^2$ for $1 \leq p \leq \infty$.
\end{abstract}

\tableofcontents

\section{Introduction and statement of results}\label{inntrodd}

Schr\"odinger operators are usually considered on the physical Hilbert space~$L^2$
of square--integrable functions. The mathematical interest in studying the $L^p$--spectrum
goes back to Barry Simon~\cite{simonLp}, who proposed to analyze localization properties of the
binding in terms of $L^p$--estimates for the eigenfunctions.
In many situations, the $L^p$--spectrum coincides with the usual $L^2$--spectrum
(see for example~\cite{hempel+voigt, sturm, daviesLp}), but there are also
examples where the spectrum does depend on~$p$ (see~\cite{hempel+voigt2, davies+simon+taylor}).

More recently, the interest in considering Schr\"odinger operators on the Banach spaces~$L^1$
and~$L^\infty$ was revived from physics due to the connection between quantum mechanics
and statistical mechanics \cite{itzykson-drouffe, isidro2, isidro1}.
In this context one is interested in the harmonic potential having
the opposite sign of the usual harmonic oscillator potential:
\begin{equation}
V({\bf r})=-{\bf r}^2,  \qquad {\bf r}=(x,y,z)\in\mathbb{R}^3.
\label{potenci}
\end{equation}
This repulsive potential might appear to be physically uninteresting. Yet it provides, among other examples, an approximate description of the expansion of the galaxies in our universe as governed by Hubble's
law~\cite{hubble, isidro4}.  Although, at first sight, this is a purely classical--mechanical problem, Eddington pointed out long ago~\cite{eddington-original} that one should also consider the corresponding scattering problem in quantum mechanics. In this paper we take up this suggestion and solve the nonrelativistic Schr\"odinger equation for the potential (\ref{potenci}) (hereafter called {\it the Hubble potential}\/, or also {\it the inverted harmonic oscillator potential}\/). We note that, in the $L^2$ case, the corresponding Hamiltonian was analyzed in
detail in~\cite{barton}, with a focus on the dynamical picture and the sojourn time. Our analysis complements that
of~\cite{barton} in that we will step outside the Hilbert space $L^2(\mathbb{R}^3)$
and consider the Banach space~$L^p(\R^3)$ for any~$p \in [1,\infty]$.
As explained in~\cite{isidro2, isidro1}, the cases~$p=1$ and~$p=\infty$ are of particular interest.

Separating variables in the standard way we may restrict our attention to the 1--dimensional problem 
\begin{equation}
{\mathcal H}\psi(x)=\lambda\psi(x),\qquad {\mathcal H}=-\frac{{\rm d}^2}{{\rm d}x^2}-x^2.
\label{rable}
\end{equation}
As we will prove later, there exist no nontrivial solutions to this eigenvalue equation that are square--integrable. Physically this easily understood, since the repulsive potential $-x^2$ possesses scattering states only, and no bound states. 

Before embarking on the resolution of~(\ref{rable}), it is instructive to put the $L^p$--{\em{spectral theory}} for the Hamiltonian (\ref{rable}) in the context
of related harmonic potentials. As is well known, the Hamitonian of the standard harmonic oscillator
\[ -\frac{{\rm d}^2}{{\rm d}x^2} + x^2 \]
on the Hilbert space $L^2(\R)$ has a purely discrete spectrum and eigenfunctions given by
\beq \label{Hermite}
\psi_n(x) = H_n(x)\, {\rm e}^{-\frac{x^2}{2}}\:,
\eeq
where the $H_n(x)$ are Hermite polynomials.
Having an exponential decay at infinity, these eigenfunctions are also in~$L^p$ for all~$p \in [1,\infty]$.
Moreover, for any~$\lambda$ not in the $L^2$--spectrum, the Green's kernel~$s_\lambda(u,u')$ can be constructed
explicitly in terms of parabolic cylinder functions (for details see~(\ref{eije}) and~(\ref{eixof}) below).
Since this Green's kernel again decays exponentially at infinity, the corresponding integral operator
defines the resolvent as a bounded operator on~$L^p$ for all~$p \in [1,\infty]$.
This shows that the spectrum of the harmonic oscillator is indeed independent of~$p$.

The paper \cite{davies+kuijlaars} is concerned with {\em{complex deformations}} of the harmonic oscillator of the form
\begin{equation} 
-\frac{{\rm d}^2}{{\rm d}x^2} + {\rm e}^{{\rm i} \varphi}\, x^2 ,\qquad \varphi \in (-\pi, \pi) \:. 
\label{Hdeform}
\end{equation}
The corresponding eigenfunctions are obtained by complex continuation of the eigensolutions
\beq \label{Hermitecc}
\psi(x) = H_n(z)\, {\rm e}^{-\frac{z^2}{2}} \qquad 
z(x) = {\rm e}^{\frac{{\rm i} \varphi}{4}}\, x\:.
\eeq
These solutions again decay exponentially at infinity and are thus in $L^p$ for all $p \in [1,\infty]$.
The point of interest in \cite{davies+kuijlaars} is that these eigenfunctions do not form a Riesz basis
in $L^p$. Constructing the resolvent via the Green's kernel, one can again show that the
spectrum of the Hamiltonians (\ref{Hdeform}) is independent of $p$.

The situation changes drastically when the angle $\varphi$ in (\ref{Hdeform}) is chosen
as $\varphi=\pi$, which gives precisely the Hubble potential (\ref{rable}).
Then the complex--deformed  eigenfunctions (\ref{Hermitecc}) become
\[ \psi(x) = H_n\big({\rm e}^{\frac{{\rm i} \pi}{4}}\:x \big)\: {\rm e}^{-\frac{{\rm i} x^2}{2}} \:. \]
Now the exponential is a phase factor which no longer gives decay at infinity.
As we shall see in this paper, this indeed gives rise to an interesting $p$-dependence of the spectrum.
Here is our main result:
\begin{Thm} \label{thmmain}
The spectrum of the Hamiltonian~${\mathcal H}$ on~$L^p(\R)$ with domain of
definition~${\mathcal D}({\mathcal H})=C^\infty_0(\R) \subset L^2(\R)$ is the strip
\beq \label{strip}
\left\{ \begin{array}{cl} \displaystyle \big| {\rm Im} \lambda \big| \leq \frac{2}{p} - 1
& {\rm if}\, p \leq 2 \\[1em]
\displaystyle \big| {\rm Im} \lambda \big| \leq 1 - \frac{2}{p}
& {\rm if}\, p > 2\:. \end{array} \right.
\eeq
In the case~$p>2$, the interior of this strip is a point spectrum.
In the case~$p=\infty$, the whole strip is in the point spectrum.
\end{Thm}

\section{Diagonalization of the Hubble Hamiltonian}\label{meka}
We consider the Hubble Hamiltonian
\begin{equation}
{\mathcal H}=-\frac{{\rm d^2}}{{\rm d}x^2}-x^2
\label{Hinv}
\end{equation}
on the Banach spaces $L^p(\mathbb{R})$, with $1\leq p\leq \infty$, with domain of definition
$$
{\mathcal D}({\mathcal H})=C^{\infty}(\mathbb{R})\cap L^p(\mathbb{R}) \:.
$$
In the case $p=2$,
Nelson's commutator theorem and the Faris--Lavine theorem imply that the Hubble
Hamiltonian (\ref{Hinv}) is essentially selfadjoint on $C^\infty_0(\R) \subset L^2(\R)$
(see \cite[Thm~X36, Thm~X38]{reed+simon2} and the Corollaries thereafter).

\subsection{Eigenfunctions}
Our starting point is the eigenvalue equation
\begin{equation}
\left(\frac{{\rm d}^2}{{\rm d}x^2}+x^2\right)\psi(x)=-\lambda \psi(x),\quad x\in\mathbb{R},\quad \lambda\in\mathbb{C}.
\label{auto}
\end{equation}
The eigenvalue $\lambda$ need not be real: Indeed, in the case~$p \neq 2$,
the notion of selfadjointness is not defined, so that the eigenvalues are not guaranteed to be real. 

A fundamental set of solutions to~(\ref{auto}), known as {\it parabolic cylinder functions}\/, can be found in the literature \cite{lebedev, special}. However, for later purposes it will be illustrative to recall the actual resolution of the differential equation (\ref{auto}). Thus we first look for a factorization of $\psi(x)$ in the form
\begin{equation}
\psi(x)=v(x)\exp(\beta x^2),\qquad \beta\in\mathbb{C},
\label{fakto}
\end{equation}
where $\beta$ is some constant to be chosen later. With (\ref{fakto}) in (\ref{auto}) one finds
\begin{equation}
\frac{{\rm d}^2}{{\rm d}x^2}v(x)+4\beta x \frac{{\rm d}}{{\rm d}x}v(x)+\left[(2\beta+\lambda)+(4\beta^2+1)x^2\right]v(x)=0.
\label{rizat}
\end{equation}
The choice $\beta={\rm i}/2$ simplifies (\ref{rizat}) to
\begin{equation}
\frac{{\rm d}^2}{{\rm d}x^2}v(x)+ 2{\rm i}x \frac{{\rm d}}{{\rm d}x}v(x)+({\rm i}+\lambda)v(x)=0.
\label{facil}
\end{equation}
{}Finally the change of variables $z={\rm e}^{{\rm i}\frac{3\pi}{4}}x$ reduces (\ref{facil}) to 
\begin{equation}
\frac{{\rm d}^2}{{\rm d}z^2}\bar v(z)-2z\frac{{\rm d}}{{\rm d}z}\bar v(z)-(1-{\rm i}\lambda)\bar v(z)=0,
\label{ident}
\end{equation}
where we have defined 
\[ \bar v(z):=v\left({\rm e}^{-{\rm i}\frac{3\pi}{4}}z\right)=v(x) \:. \]
Now~(\ref{ident}) is a particular case of the Hermite differential equation on the complex
plane~\cite[Chapter~10]{lebedev}, 
\begin{equation}
\frac{{\rm d}^2}{{\rm d}z^2}f(z)-2z\frac{{\rm d}}{{\rm d}z}f(z)+2\nu f(z)=0 \:, \qquad \nu\in\mathbb{C}\:,
\label{ermitico}
\end{equation}
where the relation between the eigenvalue $\lambda$ and the index $\nu$ is
\begin{equation}
\quad\lambda=-{\rm i}\,(2\nu+1) \:.
\label{bezz}
\end{equation}
It turns out that for~$\nu \in \C \setminus \N$, the Hermite functions (see~\cite[eq.~(10.4.3)]{lebedev}
\[ H_{\nu}(z)=\frac{1}{2\Gamma(-\nu)}\sum_{n=0}^{\infty}\frac{(-1)^n}{n!}\Gamma\left(\frac{n-\nu}{2}\right)(2z)^{n} \:, \]
solves~(\ref{ermitico}). Now $H_{\nu}(z)$ defines an entire function of $z\in\mathbb{C}$.
In the limit when~$\nu$ tends to an integer, the above series reduces to the
corresponding Hermite polynomials in~\eqref{Hermite}.
Moreover, the general solution to the Hermite equation (\ref{ermitico}) can be expressed as a linear combination of the two functions $H_{\nu}(z)$ and $\exp(z^2) \,H_{-(\nu+1)}({\rm i}z)$, since both satisfy (\ref{ermitico}) and have a nonvanishing Wronskian. Thus the general solution to~(\ref{ident}) is a linear combination of the two linearly independent functions
\[ \bar v_{\nu}(z)=H_{\nu}(z),\qquad
\bar v_{-(\nu+1)}(z)=\exp(z^2)H_{-(\nu+1)}({\rm i}z), \]
and the general eigenfunction (\ref{fakto}) becomes a linear combination of the two linearly independent eigenfunctions
\begin{align}
\psi_{\nu}(x) &:=\exp\left(\frac{{\rm i}}{2}x^2\right)H_{\nu}\left({\rm e}^{{\rm i}\frac{3\pi}{4}}x\right),
\label{eije}\\
\psi_{-(\nu+1)}(x) &:= \exp\left(-\frac{{\rm i}}{2}x^2\right)H_{-(\nu+1)}\left({\rm e}^{{\rm i}\frac{5\pi}{4}}x\right),
\label{eixof}
\end{align}
both corresponding to the same eigenvalue $\lambda$ in~\eqref{bezz}.

\subsection{Asymptotics of Eigenfunctions}
We now determine the asymptotics of the solutions~\eqref{eije} and~\eqref{eixof} as~$x \rightarrow \pm \infty$.
We begin with the function~$\psi_{\nu}$. Asymptotically as~$x \rightarrow \infty$,
the argument of the Hermite functions~$H_{\nu}(z)$ lies on the ray $z={\rm e}^{{\rm i}\frac{3\pi}{4}}x$
with~$x>0$. The asymptotics of the Hermite functions is given by \cite[eq.~(10.6.7)]{lebedev}
\[ H_{\nu}(z)\sim (2z)^{\nu}-\frac{\sqrt{\pi}{\rm e}^{{\rm i}\pi\nu}}{\Gamma(-\nu)}\,z^{-\nu-1}\,{\rm e}^{z^2},\quad 
\vert z\vert\to\infty,\quad \frac{\pi}{4}<\arg(z)< \frac{5 \pi}{4}\:. \]
We thus obtain
\[ H_{\nu}\left({\rm e}^{{\rm i}\frac{3\pi}{4}}x\right)\sim a_\nu \,x^{\nu}+b_{\nu} \,x^{-(\nu+1)}\: {\rm e}^{-{\rm i}x^2}
\qquad \text{as~$x\to\infty$} \:, \]
where the coefficients given by
\[ a_{\nu}=2^{\nu}\, {\rm e}^{{\rm i}\frac{3\nu\pi}{4}}, \quad b_{\nu}=-\frac{\sqrt{\pi}\:
{\rm e}^{{\rm i}\pi\nu}}{\Gamma(-\nu)}\:
{\rm e}^{-{\rm i}(\nu+1)\frac{3\pi}{4}} \:. \]
As a consequence, the wavefunction (\ref{eije}) has the asymptotic behavior
\begin{equation}
\psi_{\nu}(x) \sim a_\nu \,x^{\nu} \,{\rm e}^{{\rm i}x^2/2}+b_{\nu} \,x^{-(\nu+1)} \,{\rm e}^{-{\rm i}x^2/2}
\qquad \text{as~$x\to\infty$} \:.
\label{zeppo}
\end{equation}
Asymptotically as~$x \rightarrow -\infty$, the argument of the Hermite functions in~\eqref{eije}
is on the ray $z={\rm e}^{{\rm i}\frac{7\pi}{4}} |x|$. 
The corresponding asymptotics is~\cite[eq.~(10.6.6)]{lebedev}
\beq \label{mikro}
H_{\nu}(z)\sim (2z)^{\nu}, \quad \vert z\vert\to\infty,\quad \vert\arg(z)\vert<\frac{3\pi}{4} \:.
\eeq
Then
\[ H_{\nu}({\rm e}^{{\rm i}\frac{7\pi}{4}} \,|x|)\sim c_{\nu} |x|^{\nu}, \quad x\to-\infty,
\quad c_{\nu}:=2^{\nu}{\rm e}^{{\rm i}\frac{7\nu\pi}{4}} \:, \]
and the wavefunction (\ref{eije}) becomes
\begin{equation}
\psi_{\nu}(x)\sim c_{\nu}\, |x|^{\nu}{\rm e}^{{\rm i}x^2/2}
\qquad \text{as~$x\to-\infty$} \:.
\label{goodman}
\end{equation}

For the asymptotics of the solution~$\psi_{-(\nu+1)}$ as~$x \rightarrow \infty$,
we make use of the asymptotic expansion of the Hermite function~\cite[eq.~(10.6.8)]{lebedev}
\[ H_{\nu}(z)\sim (2z)^{\nu}-\frac{\sqrt{\pi}\,{\rm e}^{-{\rm i}\pi\nu}}{\Gamma(-\nu)}\:z^{-\nu-1}\,{\rm e}^{z^2},\quad 
\vert z\vert\to\infty,\quad -\frac{5\pi}{4} <\arg(z)< -\frac{\pi}{4}\:, \]
and the replacement $\nu\to-(\nu+1)$ leads to
\[ H_{-(\nu+1)}(z)\sim (2z)^{-(\nu+1)}-\frac{\sqrt{\pi}\,{\rm e}^{{\rm i}\pi(\nu+1)}}{\Gamma(\nu+1)}\,z^{\nu}\,{\rm e}^{z^2},
\quad -\frac{5\pi}{4} <\arg(z)< -\frac{\pi}{4}\:. \]
Hence
\[ H_{-(\nu+1)}\left({\rm e}^{{\rm i}\frac{5\pi}{4}}x\right)\sim d_{\nu} \,x^{-(\nu+1)}+e_{\nu} \,x^{\nu}\,{\rm e}^{{\rm i}x^2},\quad x\to\infty \:, \]
where the $x$--independent coefficients $d_{\nu}$ and $e_{\nu}$ are
\[ d_{\nu}=2^{-(\nu+1)}\,{\rm e}^{-{\rm i} \frac{5\pi}{4}(\nu+1)}, \quad e_{\nu}=-\frac{\sqrt{\pi}\,{\rm e}^{{\rm i}\pi(\nu+1)}}{\Gamma(\nu+1)}\: {\rm e}^{{\rm i}\frac{5\pi}{4}\nu} \:. \]
Thus the wavefunction (\ref{eixof}) reads asymptotically,
\begin{equation}
\psi_{-(\nu+1)}(x)\sim d_{\nu}\, x^{-(\nu+1)}\,{\rm e}^{-{\rm i}x^2/2}+e_{\nu}\, x^{\nu}\,{\rm e}^{{\rm i}x^2/2}
\qquad \text{as~$x\to \infty$} \:. \label{burro}
\end{equation}
Finally, for the asymptotics~$x \rightarrow -\infty$, we again use~(\ref{mikro})
for~$\nu$ replaced by~$-(\nu+1)$,
\[ H_{-(\nu+1)}(z)\sim (2z)^{-(\nu+1)}, \quad \vert z\vert\to\infty,\quad \vert\arg(z)\vert<\frac{3\pi}{4} \:. \]
Hence
\[ H_{-(\nu+1)}\left({\rm e}^{{\rm i}\frac{\pi}{4}}\,|x| \right)\sim f_{\nu} \,|x|^{-(\nu+1)}, 
\quad f_{\nu}:=2^{-(\nu+1)}\: {\rm e}^{-{\rm i} \frac{(\nu+1)\pi}{4}}, \quad x\to-\infty \:. \]
We conclude that the wavefunction (\ref{eixof}) has the asymptotics
\begin{equation}
\psi_{-(\nu+1)}(x)\sim f_{\nu}\,|x|^{-(\nu+1)} \,{\rm e}^{-{\rm i}x^2/2}
\qquad \text{as~$x\to-\infty$} \:.
\label{goodmanbb}
\end{equation}

\section{Resolvent Estimates}
We now write the eigenvalue equation~\eqref{auto} as the one-dimensional
Schr\"odinger equation
\beq \label{SLE}
{\mathcal{H}} \phi(x) = \lambda \, \phi(x) \qquad \text{with} \quad
{\mathcal{H}} = -\frac{\rm{d}^2}{{\rm{d}}x^2} - x^2 \:.
\eeq
We may assume that~$\lambda$ is in the upper half plane,
\beq \label{imlam}
\im \lambda \geq 0 \:,
\eeq
because otherwise, we may analyze the complex conjugate of~\eqref{SLE}.
As~$x \rightarrow \pm \infty$, the solutions have the asymptotics
\beq \label{asy0}
\phi(x) = c^\pm_1\: {\rm{e}}^{\frac{{\rm{i}} x^2}{2} +\frac{{\rm{i}} \lambda-1}{2}\, \log |x|} \:\Big(1 + \O\big( |x|^{-\frac{1}{2}} \big) \Big) 
+ c^\pm_2\: {\rm{e}}^{-\frac{{\rm{i}} x^2}{2} - \frac{{\rm{i}} \lambda+1}{2}\, \log |x|} \:\Big(1 + \O\big( |x|^{-\frac{1}{2}} \big) \Big) \:.
\eeq
This asymptotics can be understood immediately by computing the second derivatives
of the approximate solution,
\begin{align*}
\phi_\text{approx}(x) &\!:= c^\pm_1\: {\rm{e}}^{\frac{{\rm{i}} x^2}{2} +\frac{{\rm{i}} \lambda-1}{2}\, \log |x|} + c^\pm_2\: {\rm{e}}^{-\frac{{\rm{i}} x^2}{2} - \frac{{\rm{i}} \lambda+1}{2}\, \log |x|} \\
\phi'_\text{approx}(x) &= c^\pm_1\: \Big( {\rm{i}}x + \frac{{\rm{i}} \lambda-1}{2x}\Big) \:{\rm{e}}^{\frac{{\rm{i}} x^2}{2} +\frac{{\rm{i}} \lambda-1}{2}\, \log |x|} \\
&\quad +c^\pm_2\: \Big( -{\rm{i}}x - \frac{{\rm{i}} \lambda+1}{2x}\Big) \:{\rm{e}}^{-\frac{{\rm{i}} x^2}{2} - \frac{{\rm{i}} \lambda+1}{2}\, \log |x|}  \\
\phi''_\text{approx}(x) &= c^\pm_1\: \bigg\{ {\rm{i}} + \Big( {\rm{i}}x + \frac{{\rm{i}} \lambda-1}{2x}\Big)^2 + \O\big(x^{-2}\big)\bigg\} \:{\rm{e}}^{\frac{{\rm{i}} x^2}{2} +\frac{{\rm{i}} \lambda-1}{2}\, \log |x|} \\
&\quad +c^\pm_2\: \bigg\{ -{\rm{i}} + \Big( -{\rm{i}}x - \frac{{\rm{i}} \lambda+1}{2x}\Big)^2 
+ \O\big(x^{-2}\big)\bigg\} \:{\rm{e}}^{-\frac{{\rm{i}} x^2}{2} - \frac{{\rm{i}} \lambda+1}{2}\, \log |x|} \\
&= c^\pm_1\: \bigg\{ -x^2 - \lambda + \O\big(x^{-2}\big) \bigg\} \:{\rm{e}}^{\frac{{\rm{i}} x^2}{2} +\frac{{\rm{i}} \lambda-1}{2}\, \log |x|} \\
&\quad +c^\pm_2\: \bigg\{ -x^2 - \lambda + \O\big(x^{-2}\big)\bigg\} \:{\rm{e}}^{-\frac{{\rm{i}} x^2}{2} - \frac{{\rm{i}} \lambda+1}{2}\, \log |x|} \\
&= \big(-x^2 - \lambda \big)\: \phi_\text{approx}(x) \Big( 1+\O\big(x^{-2}\big) \Big)\:.
\end{align*}

We introduce the functions~$\acute{\phi}$ and~$\grave{\phi}$ as solutions
with the fastest possible asymptotic decay, i.e.\
\begin{align}
\acute{\phi}(x) &= {\rm{e}}^{\frac{{\rm{i}} x^2}{2} +\frac{{\rm{i}} \lambda-1}{2}\, \log |x|} \:\Big(1 + \O\big( (-x)^{-\frac{1}{2}} \big) \Big) 
&&\text{as~$x \rightarrow -\infty$} \label{asy1} \\
\grave{\phi}(x) &= {\rm{e}}^{\frac{{\rm{i}} x^2}{2} +\frac{{\rm{i}} \lambda-1}{2}\, \log |x|} \:\Big(1 + \O\big( x^{-\frac{1}{2}} \big) \Big) 
&&\text{as~$x \rightarrow +\infty$} \:. \label{asygravephi}
\end{align}

\begin{Lemma} The solutions~$\acute{\phi}$ and~$\grave{\phi}$ form a fundamental system.
\end{Lemma}
\Proof Comparing~\eqref{asy1} with~\eqref{goodman} and using~\eqref{bezz}, one concludes
that~$\acute{\phi}$ is a multiple of~$\psi_\nu$.
According to~\eqref{zeppo} and noting that~$b_\nu \neq 0$
(here we make use of the fact that~$\im \lambda \geq 0$, implying that~$\nu \not \in \N_0$), one sees
the asymptotics~$\acute{\phi}$ as~$x \rightarrow \infty$ is different from the asymptotics
of~$\grave{\phi}$ in~\eqref{asygravephi}. Hence~$\acute{\phi}$ and~$\grave{\phi}$ are linearly
independent.
\QED

The fundamental solutions~$\acute{\phi}$ and~$\grave{\phi}$ have the asymptotics
\begin{align}
\acute{\phi}(x) &= \acute{c}_1\: {\rm{e}}^{\frac{{\rm{i}} x^2}{2} +\frac{{\rm{i}} \lambda-1}{2}\, \log |x|} \:\Big(1 + \O\big( x^{-\frac{1}{2}} \big)
 \Big) \notag \\
 &\quad + \acute{c}_2\: {\rm{e}}^{-\frac{{\rm{i}} x^2}{2} - \frac{{\rm{i}} \lambda+1}{2}\, \log |x|} \:\Big(1 + \O\big( x^{-\frac{1}{2}} \big) \Big) 
&&\text{as~$x \rightarrow +\infty$} \\
\grave{\phi}(x) &= \grave{c}_1\: {\rm{e}}^{\frac{{\rm{i}} x^2}{2} +\frac{{\rm{i}} \lambda-1}{2}\, \log |x|}
\:\Big(1 + \O\big( (-x)^{-\frac{1}{2}} \big) \Big) \notag \\
&\quad+ \grave{c}_2\: {\rm{e}}^{-\frac{{\rm{i}} x^2}{2} - \frac{{\rm{i}} \lambda+1}{2}\, \log |x|} \:\Big(1 + \O\big( (-x)^{-\frac{1}{2}} \big) \Big)
&&\text{as~$x \rightarrow -\infty$}  \:. \label{asy4}
\end{align}
Here the parameters~$\acute{c}_2$ and~$\grave{c}_2$ are both non-zero,
because otherwise the solutions~$\acute{\phi}$ and~$\grave{\phi}$ would be linearly
dependent in view of~\eqref{asy1} and~\eqref{asygravephi}.
Computing their Wronskian at for example $x=+\infty$ gives
\begin{align}
&w(\acute{\phi}, \grave{\phi}) = \lim_{x \rightarrow \infty}
\big( \acute{\phi}'(x)\: \grave{\phi}(x) - \acute{\phi}(x)\: \grave{\phi}'(x) \big) \notag \\
&= \acute{c}_2\: \lim_{x \rightarrow \infty}
w \big( {\rm{e}}^{-\frac{{\rm{i}} x^2}{2} - \frac{{\rm{i}} \lambda+1}{2}\, \log x}, \;
{\rm{e}}^{\frac{{\rm{i}} x^2}{2} +\frac{{\rm{i}} \lambda-1}{2}\, \log x} \big) \notag \\
&= \acute{c}_2\: \lim_{x \rightarrow \infty}
\bigg\{ \Big( -{\rm{i}}x - \frac{{\rm{i}} \lambda+1}{2x} \Big) - \Big( {\rm{i}}x + \frac{{\rm{i}} \lambda-1}{2x}\Big) \bigg\}\;
{\rm{e}}^{-\frac{{\rm{i}} x^2}{2} - \frac{{\rm{i}} \lambda+1}{2}\, \log x}\:
{\rm{e}}^{\frac{{\rm{i}} x^2}{2} +\frac{{\rm{i}} \lambda-1}{2}\, \log x} \notag \\
&= \acute{c}_2\: \lim_{x \rightarrow \infty}
\Big\{ -2{\rm{i}}x - \frac{{\rm{i}} \lambda}{x}\Big\}\;
{\rm{e}}^{-\log x} = -2 {\rm{i}} \acute{c}_2 \neq 0\:. \label{wval}
\end{align}
Since this Wronskian is non-zero, we may introduce the {\em{Green's kernel}} by
\beq \label{sldef}
s_\lambda(x,x') = \frac{1}{w(\acute{\phi}, \grave{\phi})} \times \left\{
\begin{aligned} \acute{\phi}(x)\: \grave{\phi}(x') &\quad&& \text{if~$x \leq x'$} \\
\acute{\phi}(x')\: \grave{\phi}(x) &&& \text{if~$x' < x$}\:.
\end{aligned}  \right.
\eeq
We also consider the Green's kernel as the integral kernel of a
corresponding operator~$s_\lambda$, i.e.
\[ (s_\lambda \psi)(x) := \int_{-\infty}^\infty s_\lambda(x,x')\: \psi(x')\: {\rm{d}}x' \:. \]
The distributional relation
\[ \big({\mathcal{H}}_x -\lambda \big)\: s_\lambda(x,x') = \delta(x-x') \]
implies that, formally, the operator~$s_\lambda$ coincides with the resolvent,
\[ s_\lambda \overset{\text{formally}}{=} \big({\mathcal{H}} -\lambda \big)^{-1}\:. \]
In order to establish the existence of the resolvent, it suffices to show that~$s_\lambda$
is a bounded operator on~$L^p$. We begin with a preparatory lemma.

\begin{Lemma} \label{lemmakerneles}
For every~$\lambda \in \C$ with~$\im \lambda \geq 0$ there is a
constant~$c=c(\lambda)$ such that
\[ \big| s_\lambda(x,x') \big| \leq \frac{c}{\sqrt{1+|x|}\: \sqrt{1+|x'|}}\;
\min \bigg( \Big| \frac{x}{x'} \Big|, \Big| \frac{x'}{x} \Big| \bigg)^{\frac{\im \lambda}{2}} \:. \]
\end{Lemma}
\Proof By symmetry, it clearly suffices to consider the case~$x'>x$, so that
\[ s_\lambda(x,x') = -\frac{\acute{\phi}(x)\: \grave{\phi}(x')}{2 {\rm{i}} \acute{c}_2}\:  \]
According to the asymptotics~\eqref{asy1}--\eqref{asy4},
\begin{align*}
\big|\acute{\phi}(x)\big| &\eqsim \frac{1}{\sqrt{|x|}}\: {\rm{e}}^{-\frac{\im \lambda}{2}\: \log|x|} \:,\;\;
\big|\grave{\phi}(x')\big| \eqsim \frac{1}{\sqrt{|x'|}}\: {\rm{e}}^{\frac{\im \lambda}{2}\: \log|x'|}
&&\text{as~$x,x' \rightarrow -\infty$} \\
\big|\acute{\phi}(x)\big| &\eqsim \frac{1}{\sqrt{|x|}}\: {\rm{e}}^{\frac{\im \lambda}{2}\: \log|x|}  \;,\;\;\;\;
\big|\grave{\phi}(x')\big| \eqsim \frac{1}{\sqrt{|x'|}}\: {\rm{e}}^{-\frac{\im \lambda}{2}\: \log|x'|}
&&\text{as~$x,x' \rightarrow +\infty$}\:.
\end{align*}
Moreover, the solutions~$\acute{\phi}$ and~$\grave{\phi}$ are clearly bounded on any compact set.
This gives the result.
\QED

\begin{Prp} \label{prpreses}
Assume that for~$p \in [1,\infty]$, the following inequality holds:
\[ \frac{\im \lambda}{2} > \bigg| \frac{1}{2}-\frac{1}{p} \bigg| \:. \]
Then the resolvent exists as a bounded operator on~$L^p(\R)$.
\end{Prp}
\Proof Let~$p, q \in [1, \infty]$ be conjugate H\"older exponents, i.e.
\beq \label{Hoelder}
\frac{1}{p} + \frac{1}{q} = 1 \:.
\eeq
Our goal is to show that there is a constant~$c=c(\lambda)$ such that
for all~$\psi \in L^p(\R)$ and~$\phi \in L^q(\R)$, the inequality
\beq \label{goal}
\int_{\R^2} \big| \overline{\phi(x)}\: s_\lambda(x,y)\: \psi(y) \big|\: {\rm{d}}x\,{\rm{d}}y 
\leq c\: \|\phi\|_{L^q}\: \|\psi\|_{L^p}
\eeq
holds. Indeed, in the case~$1 \leq p < \infty$, we can then use the fact that~$(L^p)^* = L^q$ to
conclude that~$s_\lambda$ is a bounded operator on~$L^p$.
In the case~$p=\infty$, we set~$\eta = s_\lambda \psi$.
For any given~$\varepsilon>0$, we can
choose a set~$\Omega$ with non-zero Lebesgue measure such that
\[ \eta|_\Omega \geq \|\eta\|_{L^\infty} - \varepsilon \qquad \text{or} \qquad
\eta|_\Omega \leq -\|\eta\|_{L^\infty} + \varepsilon \:. \]
Choosing~$\phi$ as the characteristic function~$\phi = \chi_{\Omega}$, it follows that
\[ \int_{\R^2} \big| \overline{\phi(x)}\: s_\lambda(x,y)\: \psi(y) \big|\: {\rm{d}}x\,{\rm{d}}y
\geq \big( \|\eta\|_{L^\infty} - \varepsilon \big) \: \|\phi\|_{L^1} \:. \]
Therefore, the inequality~\eqref{goal} implies that
\[ \big( \|\eta\|_{L^\infty} - \varepsilon \big) \leq c\: \|\psi\|_{L^\infty} \:. \]
Since~$\varepsilon$ is arbitrary, we conclude that~$s_\lambda$ is a bounded operator on~$L^\infty$.

It remains to derive the inequality~\eqref{goal}.
Employing the estimate of Lemma~\ref{lemmakerneles}, we obtain
\begin{align*}
\Big| &\int_{R^2} \overline{\phi(x)}\: s_\lambda(x,y)\: \psi(y)\: {\rm{d}}x\,{\rm{d}}y \Big| \\
& \leq c \int_{\R^2}  \frac{|\phi(x)|\: |\psi(y)|}{\sqrt{1+|x|}\: \sqrt{1+|y|}}\;
\min \bigg( \Big| \frac{x}{y} \Big|, \Big| \frac{y}{x} \Big| \bigg)^{\frac{|\im \lambda|}{2}} \: {\rm{d}}x\, {\rm{d}}y \:.
\end{align*}
Now it is useful to choose polar coordinates~$r \in \R^+$, $\varphi \in [0, 2 \pi)$, i.e.
\[ x = r \cos \varphi \:,\qquad y = r \sin \varphi \:. \]
We thus obtain the estimate
\begin{align}
\bigg| &\int_{\R^2} \overline{\phi(x)}\: s_\lambda(x,y)\: \psi(y)\: {\rm{d}}x\,{\rm{d}}y \bigg| \notag \\
& \leq c \int_0^{2 \pi} {\rm{d}}\varphi \int_0^\infty r\: {\rm{d}}r\; \frac{\big|\phi(r \cos \varphi) \big|\:
\big| \psi(r \sin \varphi) \big|}{\sqrt{1+r \,|\cos \varphi|}\: \sqrt{1+r\, |\sin \varphi|}}\;
\min \Big( \big| \cot \varphi \big|, \big| \tan \varphi \big| \Big)^{\frac{|\im \lambda|}{2}} \notag \\
& \leq c \int_0^{2 \pi} {\rm{d}}\varphi \int_0^\infty {\rm{d}}r\; \frac{\big| \phi(r \cos \varphi) \big|\:
\big| \psi(r \sin \varphi) \big|}{\sqrt{ |\cos \varphi\:
\sin \varphi|}}\; \min \Big( \big| \cot \varphi \big|, \big| \tan \varphi \big| \Big)^{\frac{|\im \lambda|}{2}} \:. \label{finin}
\end{align}
Now the $r$-integral can be estimated with H\"older's inequality,
\[ \int_0^\infty \big|\phi(r \cos \varphi)\big|\: \big|\psi(r \sin \varphi) \big|\: {\rm{d}}r
\leq \big\|\phi(.\: \cos \varphi)\big\|_{L^q}\: \big\|\psi(.\: \sin \varphi)\big\|_{L^p} \:. \]
The obtained norms can be simplified by rescaling. Namely, in the case~$1 \leq p < \infty$,
the transformation
\[ \big\| \psi(.\: \sin \varphi) \big\|_{L^p}
\leq \left( \int_{-\infty}^\infty \big|\psi(x \sin \varphi)\big|^p\: {\rm{d}}x \right)^{\frac{1}{p}} \\
= \left( \int_{-\infty}^\infty \big|\psi(u)\big|^p\: \frac{{\rm{d}}u}{|\sin \varphi|} \right)^{\frac{1}{p}} \]
implies that
\[ \big\| \psi(.\: \sin \varphi) \big\|_{L^p} \leq |\sin \varphi|^{-\frac{1}{p}}\: \|\psi\|_{L^p} \:. \]
In the case~$p=\infty$, this inequality is again satisfied trivially. Rescaling the $q$-norm
similarly, we obtain the inequality
\[ \int_0^\infty \big|\phi(r \cos \varphi) \big|\: \big|\psi(r \sin \varphi) \big|\: {\rm{d}}r
\leq |\cos \varphi|^{-\frac{1}{q}}\: |\sin \varphi|^{-\frac{1}{p}}\: \|\phi\|_{L^q}\: \|\psi\|_{L^p} \:. \]
Using this inequality in~\eqref{finin}, we get the estimate
\begin{align*}
\bigg| &\int_{\R^2} \overline{\phi(x)}\: s_\lambda(x,y)\: \psi(y)\: {\rm{d}}x\,{\rm{d}}y \bigg| \\
&\leq \|\phi\|_{L^q}\: \|\psi\|_{L^p} \int_0^{2 \pi}
\frac{1}{|\cos \varphi|^{\frac{1}{2}+\frac{1}{q}}\: |\sin \varphi|^{\frac{1}{2}+\frac{1}{p}}}\;
\min \Big( \big| \cot \varphi \big|, \big| \tan \varphi \big| \Big)^{\frac{|\im \lambda|}{2}} \: {\rm{d}}\varphi 
\end{align*}
It remains to analyze whether the $\varphi$-integral is finite. To this end, we need to
consider the poles of the integrand,
\begin{align*}
&\frac{1}{|\cos \varphi|^{\frac{1}{2}+\frac{1}{q}}\: |\sin \varphi|^{\frac{1}{2}+\frac{1}{p}}}\;
\min \Big( \big| \cot \varphi \big|, \big| \tan \varphi \big| \Big)^{\frac{|\im \lambda|}{2}} \\
&\quad
\sim \left\{ \begin{array}{cl} \displaystyle
|\sin \varphi|^{\frac{|\im \lambda|}{2}-\frac{1}{2}-\frac{1}{p}}& \text{near~$\varphi=0, \pi$} \\[0.5em]
\displaystyle
|\cos \varphi|^{\frac{|\im \lambda|}{2}-\frac{1}{2}-\frac{1}{q}} & \text{near~$\varphi=\frac{\pi}{2}, \frac{3\pi}{2}$} \\
\end{array} \right.
\end{align*}
These poles are integrable if and only if
\[ \frac{|\im \lambda|}{2}-\frac{1}{2}-\frac{1}{p} > -1 \qquad \text{and} \qquad
\frac{|\im \lambda|}{2}-\frac{1}{2}-\frac{1}{q} > -1 \:. \]
This gives the result.
\QED

\section{Localizing the Point Spectrum in the Case~$p > 2$}
\begin{Lemma} \label{lemmapg2} In the case~$p>2$, the strip
\beq \label{in1}
0 < \im \lambda < 1 - \frac{2}{p}
\eeq
is in the point spectrum. In the case~$p=\infty$, the point spectrum is given by the strip
\beq \label{in2}
0 \leq \im \lambda \leq 1 \:.
\eeq
\end{Lemma}
\Proof According to the asymptotic expansions~\eqref{asy1}--\eqref{asy4},
the fundamental solutions are bounded by
\beq \label{phies}
|\phi(x)| \lesssim c(\lambda)\;(1+|x|)^{\frac{|\im \lambda|}{2}-\frac{1}{2}} \:.
\eeq
In the case~$1 \leq p < \infty$, this function is in the Banach space~$L^p$ if and only if
\[ p \left(\frac{|\im \lambda|}{2}-\frac{1}{2} \right) < -1 \:. \]
In particular, the function~$\phi$ is in the point spectrum if~\eqref{in1} holds.

In the case~$p=\infty$, the estimate~\eqref{phies} shows that under the assumption~\eqref{in2}
the function~$\phi$ is essentially bounded and is thus an eigenvector. This concludes the proof.
\QED

\section{Localizing the Spectrum in the Case~$1 \leq p < 2$}
\begin{Lemma} \label{lemmaps2}
In the case~$1 \leq p<2$, the spectrum of the Hamiltonian contains the strip
\beq \label{in1alt}
0 < \im \lambda < \frac{2}{p} - 1\:.
\eeq
\end{Lemma}
\Proof We proceed indirectly. Assume that~$\lambda$ is not in the spectrum.
Then the resolvent~$({\mathcal{H}}-\lambda)^{-1} : L^p(\R) \rightarrow L^p(\R)$ exists and is continuous.
Given a test function~$\eta \in C^\infty_0(\R)$, the function~$\phi := ({\mathcal{H}}-\lambda)^{-1}) \eta
\in L^p(\R)$ is a weak solution of the differential equation~$({\mathcal{H}}-\lambda) \phi = \eta$.
As a consequence, outside the support of~$\eta$, the function~$\phi$ is a solution of
the ODE~\eqref{SLE}. Combining the asymptotics~\eqref{asy0} with the fact that~$\phi \in L^p(\R)$
implies that~$\phi$ must vanish identically outside the support of~$\eta$
(note that {\em{both}} summands in~\eqref{asy0} are {\em{not}} in $L^p$).
Considering the limiting case where~$\eta$ approaches a Dirac distribution, we obtain a contradiction.
\QED

\section{Proof of the Main Theorem}

\Proof[Proof of Theorem~\ref{thmmain}.]
We may assume that~$\im \lambda \geq 0$, because otherwise we may take the complex conjugate
of the equation~\eqref{SLE}.
From Proposition~\ref{prpreses}, we know that the spectrum lies inside the strip
\[ \frac{\im \lambda}{2} \leq \bigg| \frac{1}{2}-\frac{1}{p} \bigg| \:. \]
According to Lemma~\ref{lemmapg2} and Lemma~\ref{lemmaps2}, the interior of this strip
belongs to the spectrum. Since the spectrum is a closed set, we obtain~\eqref{strip}.
The statement on the point spectrum is proved in Lemma~\ref{lemmapg2}.
\QED

\Thanks {{\em{Acknowledgments:}}
We would like to thank Dirk Deckert, Martin Oelker and Peter Pickl for helpful discussions.
J.M.I.\ acknowledges support by grant no.\ ENE2015-71333-R (Spain).

\providecommand{\bysame}{\leavevmode\hbox to3em{\hrulefill}\thinspace}
\providecommand{\MR}{\relax\ifhmode\unskip\space\fi MR }
\providecommand{\MRhref}[2]{%
  \href{http://www.ams.org/mathscinet-getitem?mr=#1}{#2}
}
\providecommand{\href}[2]{#2}

\end{document}